\documentclass[a4paper]{cas-sc}

\usepackage[numbers,sort&compress]{natbib}
\usepackage{subfigure}
\usepackage{lineno}
\def\tsc#1{\csdef{#1}{\textsc{\lowercase{#1}}\xspace}}
\tsc{WGM}
\tsc{QE}

\begin{document}
\let\WriteBookmarks\relax
\def\floatpagepagefraction{1}
\def\textpagefraction{.001}

\shorttitle{}    

\shortauthors{}  

\title [mode = title]{Behavioral-Level Simulation of Digital Readout for COFFEE at LHCb Upstream Pixel Tracker}   

\author[1,2]{Xiaoxu Zhang}
\author[2]{Yang Zhou}
\cormark[1]
\author[3]{Xiaomin Wei}
\author[4]{Anqi Wang}
\author[5,2]{Leyi Li}
\author[3]{Yu Zhao}
\author[3]{Zexuan Zhao}
\author[3]{Huimin Wu}
\author[2,4]{Mingjie Feng}
\author[1]{Lei Zhang}
\author[2,6]{Jianchun Wang}
\author[2,6]{Yiming Li}

\affiliation[1]{organization={Nanjing University},
	addressline={22 Hankou Road}, 
	city={Nanjing},
	postcode={210093}, 
	country={China}}
\affiliation[2]{organization={Institute of High Energy Physics(IHEP), Chinese Academy of Sciences},
	addressline={19B Yuquan Road}, 
	city={Beijing},
	postcode={100049}, 
	country={China}}
\affiliation[3]{organization={Northwestern Polytechnical University},
	addressline={1 Dongxiang Road}, 
	city={Xi'an},
	postcode={710129}, 
	country={China}}
\affiliation[4]{organization={University of Chinese Academy of Sciences(UCAS)},
	addressline={19B Yuquan Road}, 
	city={Beijing},
	postcode={100049}, 
	country={China}}
\affiliation[5]{organization={Shandong University},
	addressline={72 Binhai Road}, 
	city={Qingdao},
	postcode={266237}, 
	country={China}}
\affiliation[6]{organization={High Energy Physics Research Center of Henan Academy of Sciences},
	addressline={228 Chongshili}, 
	city={Zhengzhou},
	postcode={450046}, 
	country={China}}

\cortext[1]{Corresponding author: Yang Zhou (zhouyang@ihep.ac.cn)}


\begin{abstract}
COFFEE series is a HVCMOS pixel sensor using the advanced 55 nm process, currently being developed for the Upstream Pixel (UP) tracker of the LHCb Upgrade II. To ensure that COFFEE will be able to handle the particle hit rates at UP tracker, which reach a maximum of 322.5 MHz/chip, detailed simulation of the digital readout circuitry was performed. Simulation results show that the column-drain readout mechanism achieves nearly 100\% efficiency when the single readout cycle does not exceed 100 ns. Meanwhile, the buffer depth and memory resources required for the peripheral readout adapted to the BXID-sharing data format are also evaluated. These provide guidance for the design of COFFEE. The column-drain readout mechanism was used in COFFEE3 (fabricated in 2025), while the peripheral readout architecture adapted to the BXID-sharing data format is implemented in CHiR (taped out in early 2026).
\end{abstract}



\maketitle

\section{Introduction}
Pioneering R\&D in HVCMOS pixel sensors \cite{9373986, Scherl_2024} using the advanced 55 nm process, the COFFEE series prototypes \cite{Li_2025, Wei2026DesignAF} are currently being developed for the Upstream Pixel tracker (UP) in the LHCb Upgrade II \cite{LHCbcollaboration:2903094}. After the LHCb Upgrade II, Run 5 will operate with proton-proton collisions at 14 TeV with 40 MHz bunch crossing (BX), and a luminosity of 1.0 $\times$ $10^{34}$ $cm^{-2}$ $s^{-1}$. The innermost chips of UP are placed 4 cm away from the beam pipe. The placement close to the beamline at such high luminosities will lead to a high hit rate, with corresponding challenges for ASICs.

To evaluate the ASIC performance, detailed simulation is essential. In the simulation, a behavioral-level model of the pixel array and peripheral readout circuits was established using SystemC, and Monte Carlo (MC) hit events were fed into the testbench. Compared to the averaged hit density value, the MC hit events exhibit fluctuations, resulting in non-uniform and bursty data traffic. This leads to more detailed and reliable results. The simulation focused on two aspects:
\begin{enumerate}
	\itemsep=0pt
	\item The pileup of the column-drain readout mechanism adopted by COFFEE3 was investigated in the high-hit-density environment of LHCb Upgrade II. As detailed in Section 3, the simulation results reveal the critical impact of the single read period on chip efficiency.
    \item Due to the bandwidth limitation of output links, the inner UP will use the bunch crossing grouping data format. Section 4 introduces the adapted peripheral readout architecture and the simulation results regarding the large memory resources required.
\end{enumerate} 
\section{Input conditions for simulation}
The row-column configuration of the pixel array and the placement of peripheral readout are shown in \textcolor{blue}{Figure 1 (a)}. The whole chip size is 2 $\times$ 2 $cm^{2}$. The sensitive area measures 1.92 $\times$ 1.8 $cm^{2}$ and contains 128 $\times$ 360 pixels. The pixel size of 150 $\times$ 50 $\mu m^{2}$ is determined by the required spatial resolution and the layout area needed for in-pixel circuitry.

The placement and orientation of the innermost chips adjacent to the beam are shown in \textcolor{blue}{Figure 1 (b)}. Two worst-case chips were selected for the following simulation.
\begin{figure*}
	\centering
	\subfigure[]{
		\includegraphics[width=0.4\columnwidth]{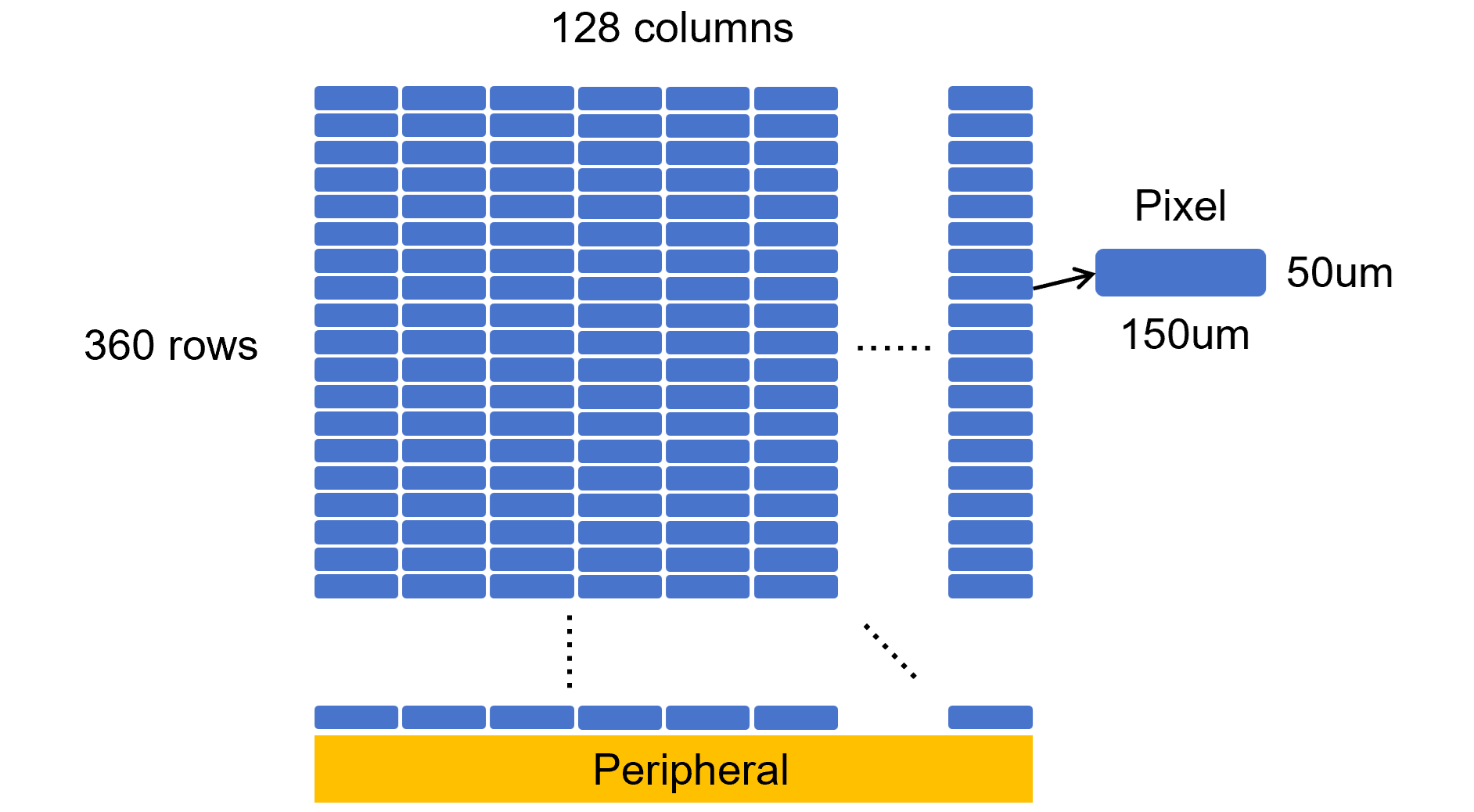}}
	\subfigure[]{
		\includegraphics[width=0.28\columnwidth]{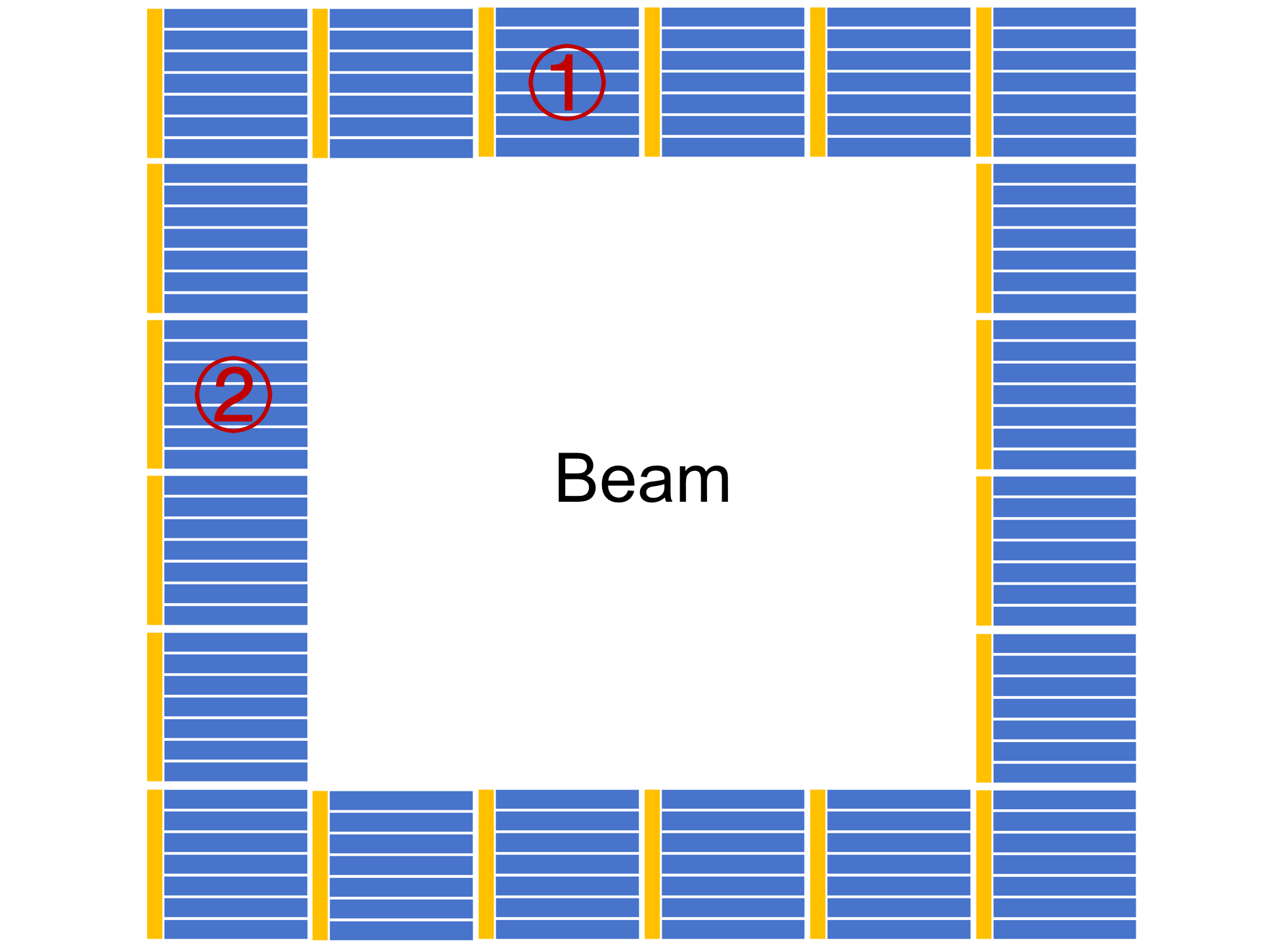}}
	\caption{(a) The row-column configuration of pixel array and the placement of peripheral readout. (b)  The placement and orientation of the innermost chips adjacent to the beam.}
	\label{FIG:1}
\end{figure*}

For MC hit events, a sample of 50,000 minimum-bias proton–proton collisions with a cluster size of 1.5 were used. 

The hit rate distributions of two selected chips are shown in \textcolor{blue}{Figure 2 (a)}. Chip 2 is the hottest, with a hit rate of 322.5 MHz, while chip 1 has a hit rate of 274.9 MHz that is distributed non-uniformly across double columns.

Also important is the hits/BXID/chip distribution, as shown in \textcolor{blue}{Figure 2 (b)}. Most values are below 30, with a long tail reaching up to 61. The peripheral readout circuits will require significant memory resources to accommodate these rare bursts. 

As shown in \textcolor{blue}{Figure 2 (c)}, a Landau distribution ranging from 200 to 1500 ns was used for the time over threshold (TOT).
\begin{figure*}
	\centering
	\subfigure[]{
		\includegraphics[width=0.32\columnwidth]{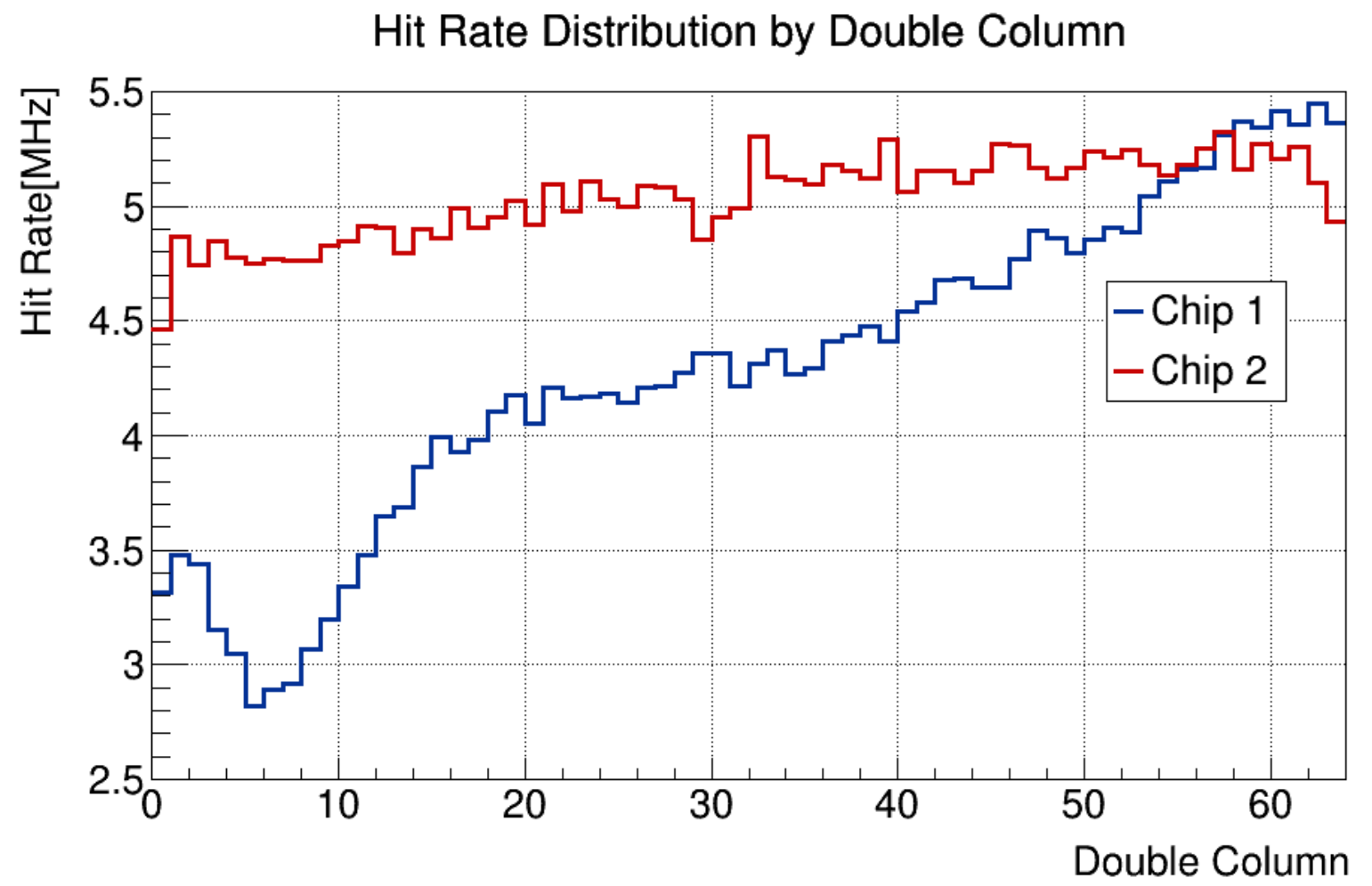}}
	\subfigure[]{
		\includegraphics[width=0.32\columnwidth]{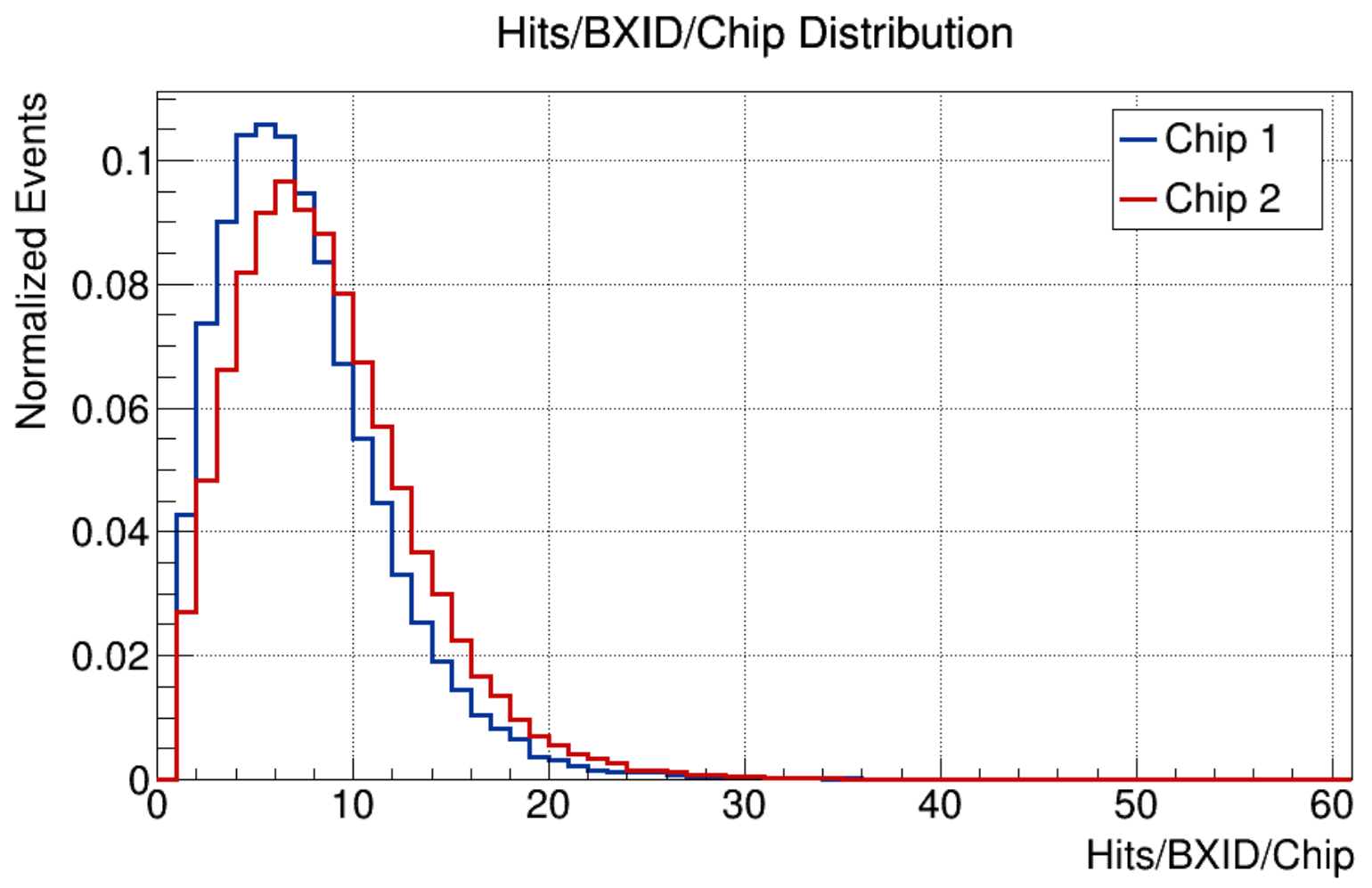}}
	\subfigure[]{
		\includegraphics[width=0.32\columnwidth]{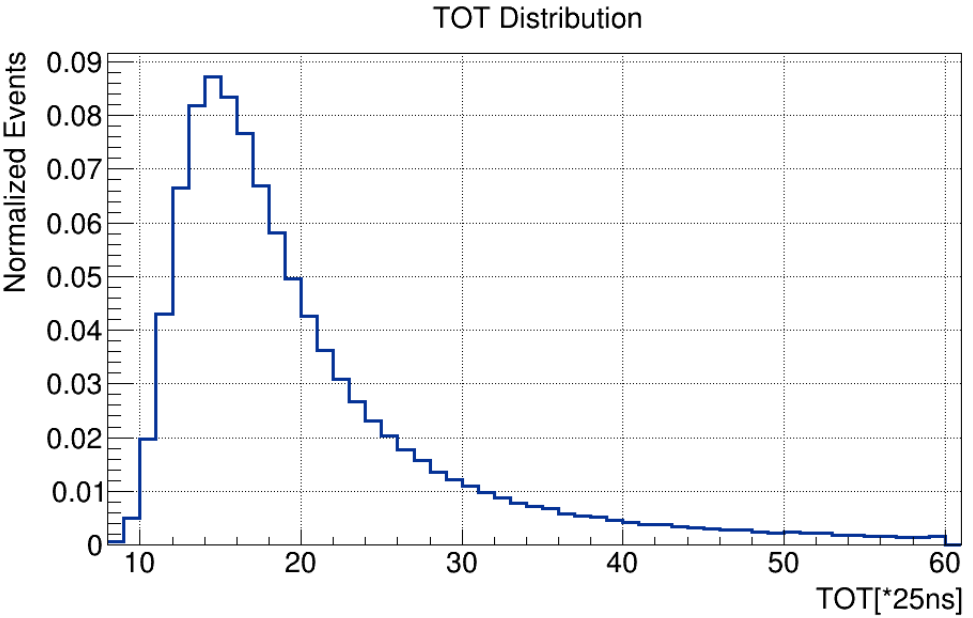}}
	\caption{The characteristics of MC hit events. (a) Hit rate distributions of two selected chips. (b) Distribution of hits/BXID for a single chip. (c) TOT distribution.}
	\label{FIG:2}
\end{figure*}

\section{Readout mechanism and chip efficiency analysis}
The column-drain readout mechanism adopted by COFFEE3  is similar to \cite{Barbero_2020}. The time stamps of the leading edge (LE) and trailing edge (TE) are stored in the in-pixel RAM cell at the end of signal pulse. The readout of pixel hits is arbitrated by a token passing scheme in each double column, with the topmost pixel having the highest priority. Each in-pixel RAM cell can store data for only one hit. New hits will be ignored until the existing data is read out, which results in efficiency loss.

The readout controller at the end of column (EoC) sends READ signals to pixels, which determine the readout cycle period for each double column. The simulation results of the chip efficiency with different READ signal widths are shown in \textcolor{blue}{Figure 3}. It can be seen that:
\begin{enumerate}
	\itemsep=0pt
	\item The chip efficiency remains near 100\% for READ signal width $\le$ 100 ns but exhibits a significant decrease for READ signal widths $>$ 100 ns.
	\item For READ signal width $>$ 100 ns, the efficiency of chip 1 not only decreases but also exhibits significant variation across different double columns. This non-uniformity in efficiency stems from the non-uniform hit rate distribution shown in \textcolor{blue}{Figure 2 (a)}, and will lead to systematic biases in the reconstructed physics quantities.
	
\end{enumerate} 
\begin{figure*}
	\centering
	\subfigure[]{
		\includegraphics[width=0.95\columnwidth]{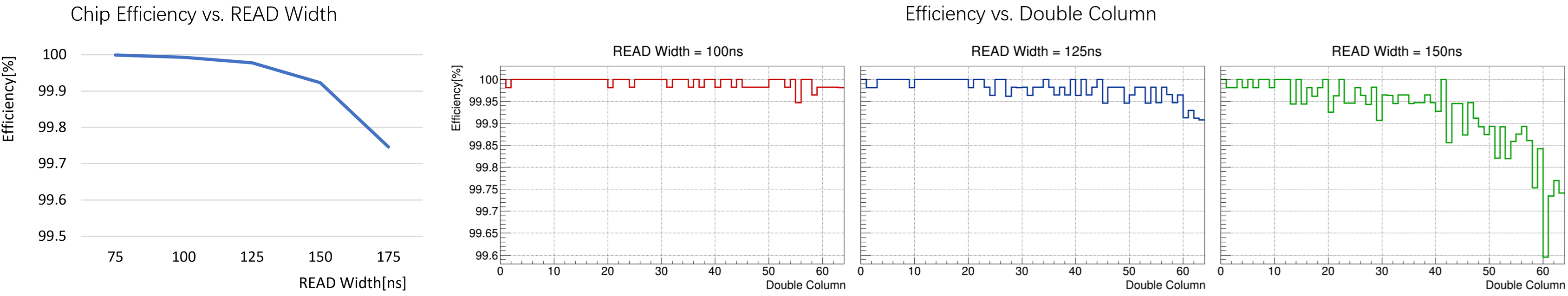}}
	\subfigure[]{
		\includegraphics[width=0.95\columnwidth]{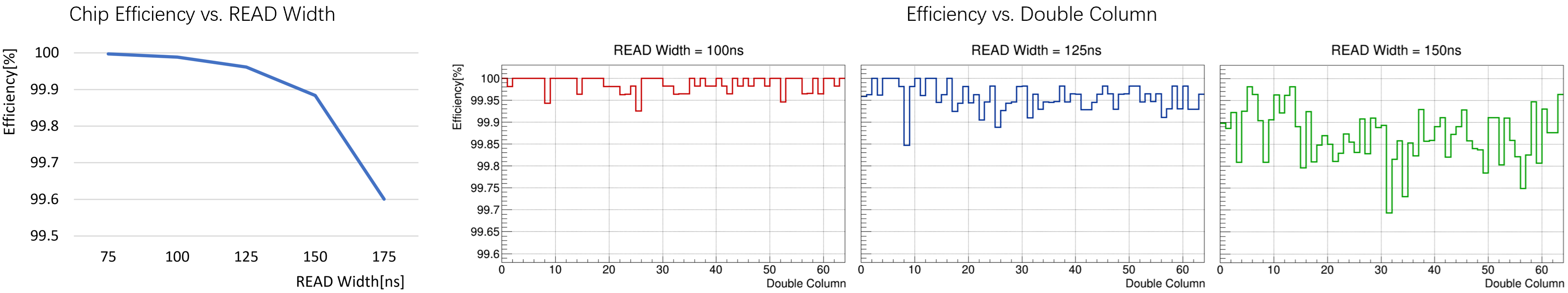}}
	\caption{The chip efficiency with different READ signal widths. (a) Chip 1. (b) Chip 2.}
	\label{FIG:3}
\end{figure*}

The READ signal width is constrained by the maximum transmission delay on the double-column data bus. A READ signal width of 100 ns demands robust bus drivers.

\section{Peripheral readout architecture adapted to the BXID-sharing data format}
\begin{figure}
	\centering
	\includegraphics[width=.5\columnwidth]{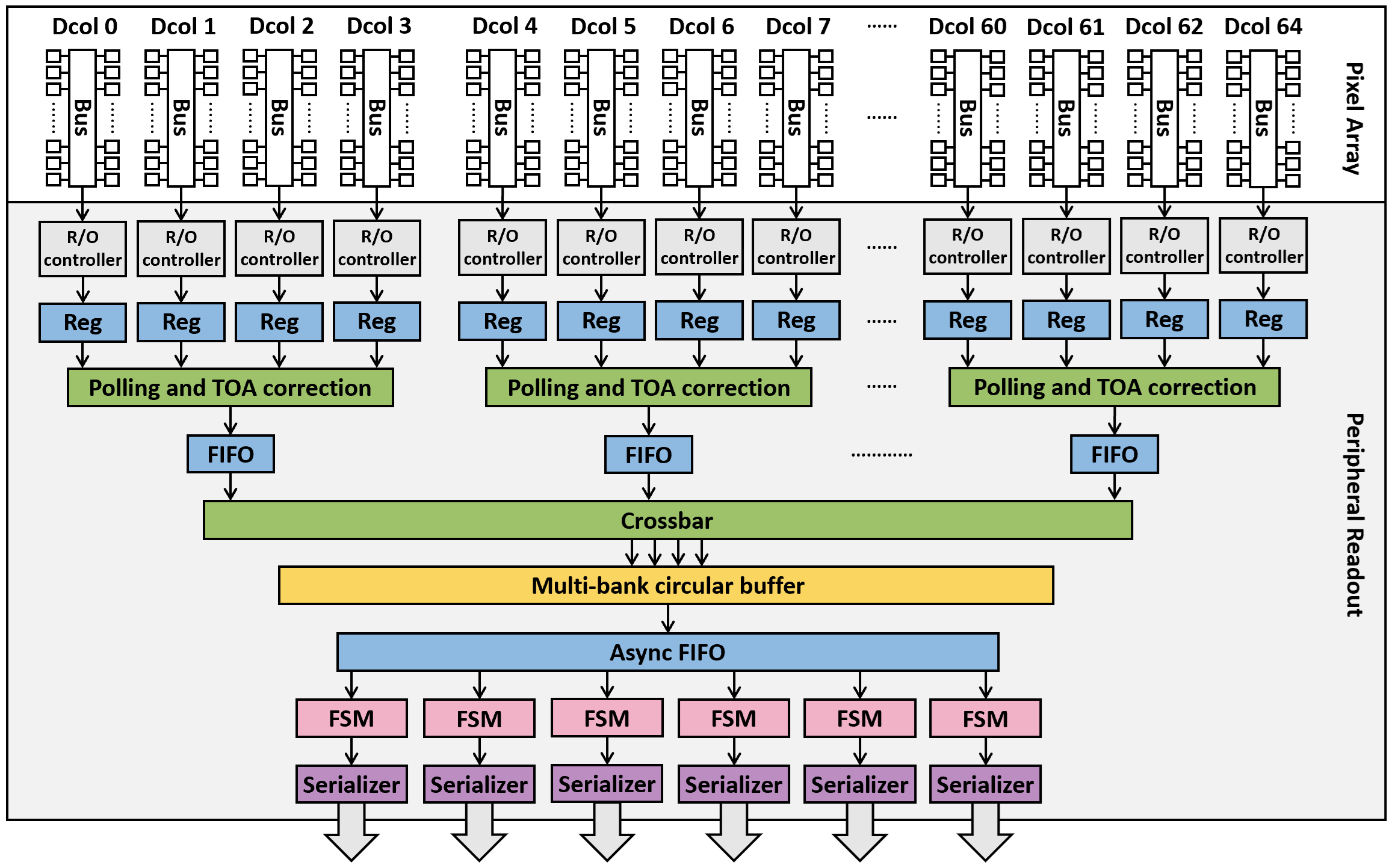}
	\caption{Block diagram of the peripheral readout architecture adapted for the BXID-sharing data format.}
	\label{FIG:4}
\end{figure}
The peripheral readout architecture adapted to the BXID-sharing data format is shown in \textcolor{blue}{Figure 4}. Each double column is equipped with an EoC readout controller, and every four readout controllers are polled for first-level data aggregation, while the TOT is used to correct the time of arrival (TOA) value via the lookup table (LUT). And then the data including TOA and address information are stored in FIFOs. The on-chip TOA correction eliminates the need to transmit the TOT bits off-chip. And the subsequent backend architecture includes a crossbar, a globally shared multi-bank circular buffer, an asynchronous FIFO, six finite state machines (FSMs) and six serializers.
\subsection{Data format and multi-bank circular buffer}
The core of the peripheral readout architecture lies in the data format. The data format used by UP during normal data taking is shown in \textcolor{blue}{Figure 5 (a)}. All detector regions that require more than one e-link/chip to transmit data will use the Normal Compact format to cope with the high hit rate. Compared with the Normal format, chip 1 and chip 2 save 39.3\% and 40.8\% of the bits, respectively, when using the Normal Compact format.

The key to realizing the BXID sharing is the globally shared multi-bank circular buffer shown in \textcolor{blue}{Figure 5 (b)}. Each bank is equipped with four independent write ports, allowing simultaneous writing of up to four data words associated with a specific TOA value. Operating at 40 MHz, in each clock cycle one bank is read out, while the TOA index of that bank is incremented by the total number of banks N to accept new data, thereby forming a circular buffer. Each bank has a width of 16 bits (7 bits for column address and 9 bits for row address) and a depth equal to the maximum number of hits/BXID/chip as shown in \textcolor{blue}{Figure 2 (b)}. 

Due to variations in TOT and the queuing during data transmission, the latency from the moment a particle hits a pixel to the time the data packet arrives at the circular buffer follows a distribution as shown in \textcolor{blue}{Figure 6 (a)}. It can be seen that the majority of latency is concentrated within 80 clock cycles, which is mainly attributed to the TOT distribution. However, a few latency values extend up to 215 clock cycles when READ signal width = 100 ns, forming a long tail in the distribution. This long tail is mainly attributed to the queuing latency in pixels. The total number of banks, N, is equal to the coverage of this latency distribution, measured in units of 40 MHz clock cycle. Truncating the long tail to save memory resources results in efficiency loss, as shown in \textcolor{blue}{Figure 6 (b)}. If the READ signal width is set to 125 ns, the long tail will become significantly longer.

For READ signal width = 100 ns, the maximum occupancy of the 16 FIFOs in the multi-bank circular buffer frontend is shown in \textcolor{blue}{Figure 6 (c)}. Since every four FIFOs correspond to an independent write port in a bank and prioritization is applied among these four FIFOs, significant differences in occupancy are observed across the FIFOs. Based on the simulation results, a FIFO depth of approximately 23 is required.
\begin{figure*}
	\centering
	\subfigure[]{
		\includegraphics[width=0.4\columnwidth]{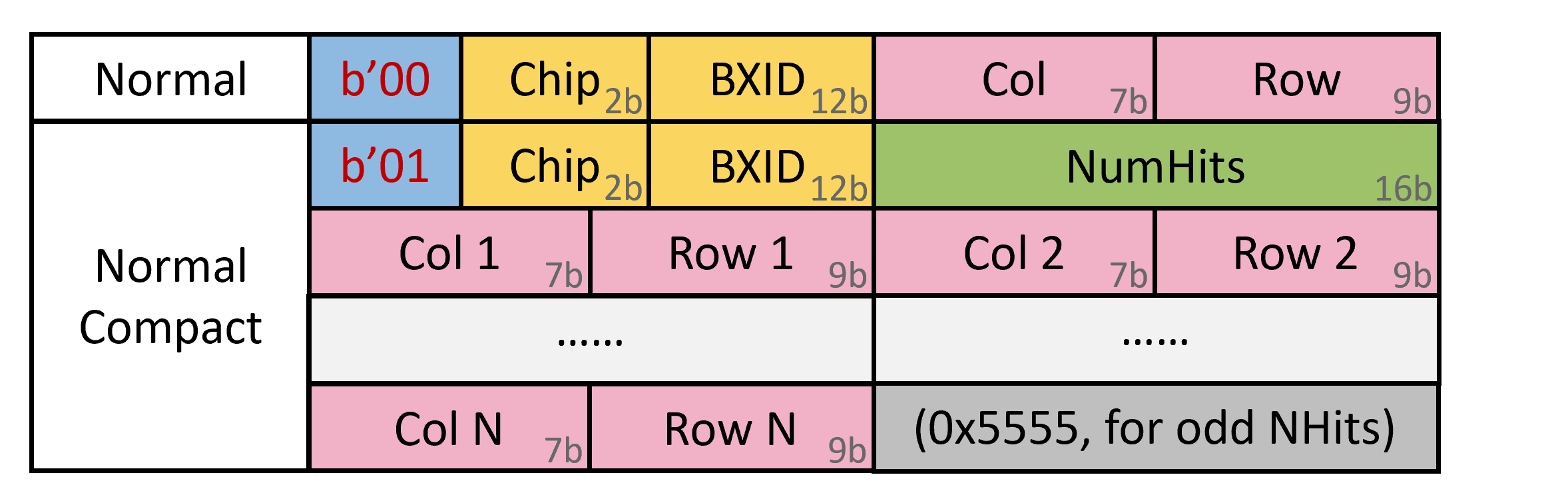}}
	\subfigure[]{
		\includegraphics[width=0.48\columnwidth]{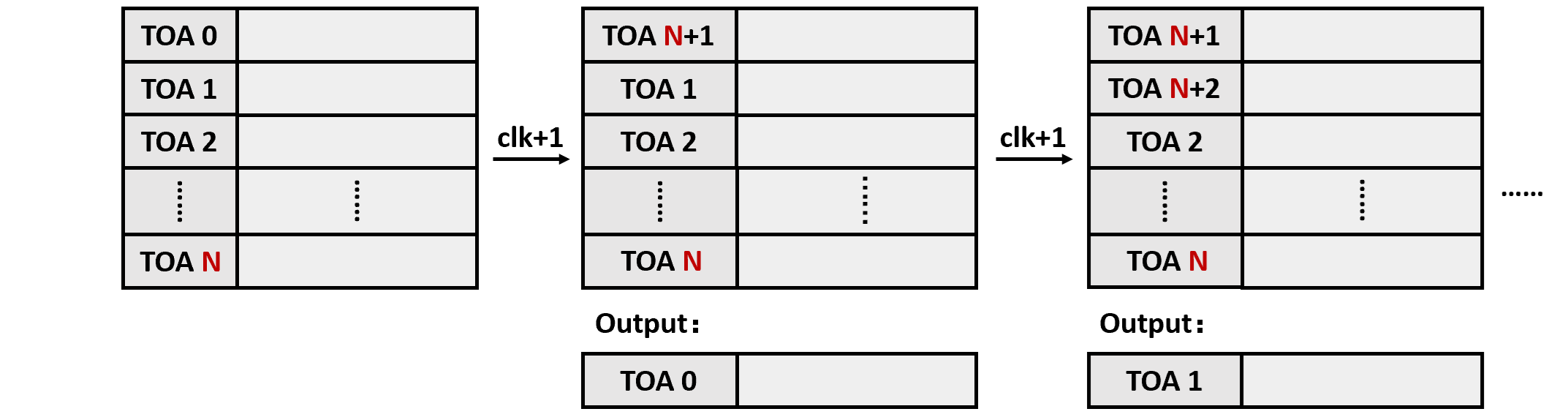}}
	\caption{(a) The Normal Compact data format used by UP. (b) Schematic diagram illustrating the mechanism of the globally shared multi-bank circular buffer.}
	\label{FIG:5}
\end{figure*}
\begin{figure*}
	\centering
	\subfigure[]{
		\includegraphics[width=0.32\columnwidth]{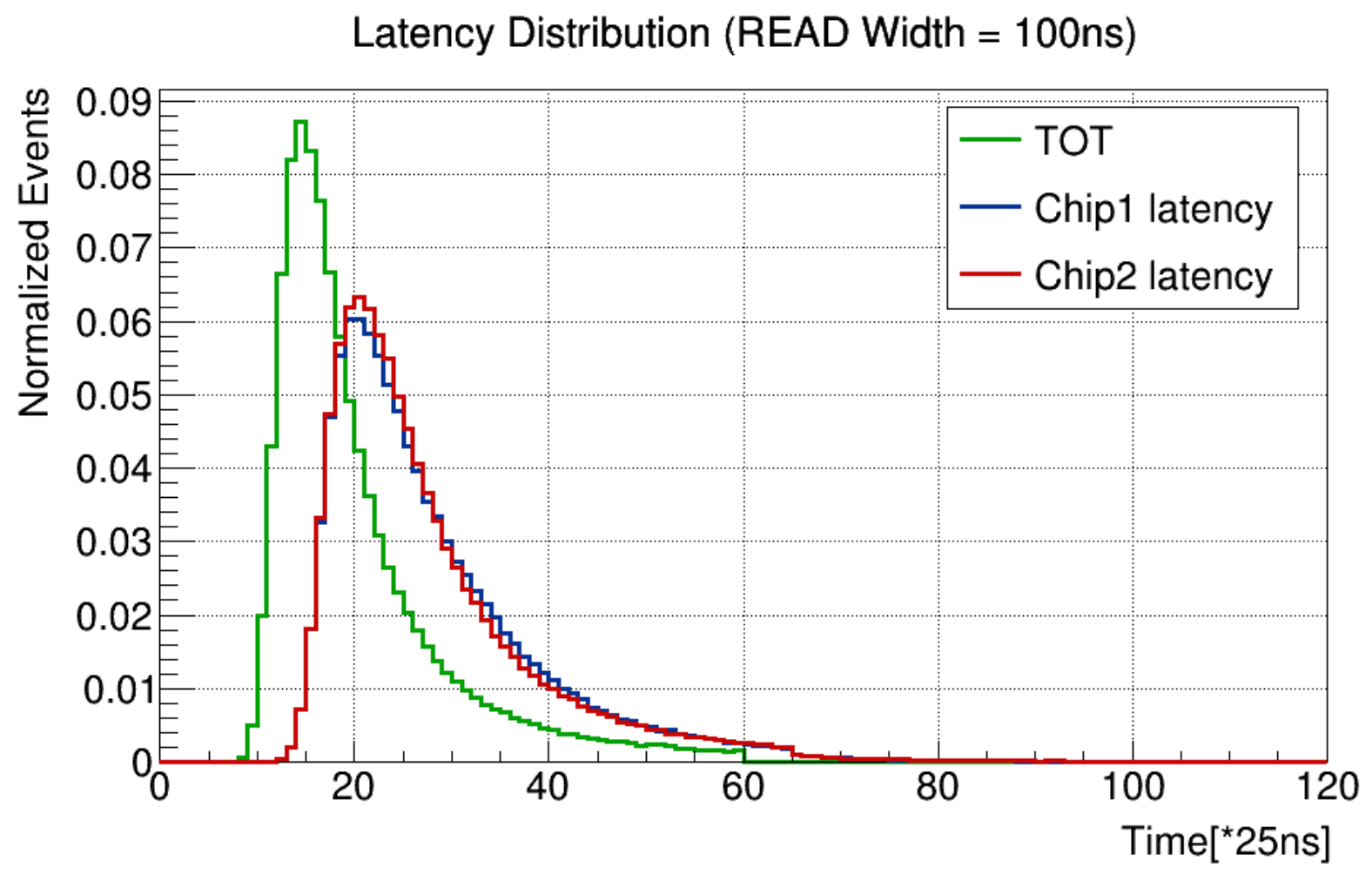}}
	\subfigure[]{
		\includegraphics[width=0.24\columnwidth]{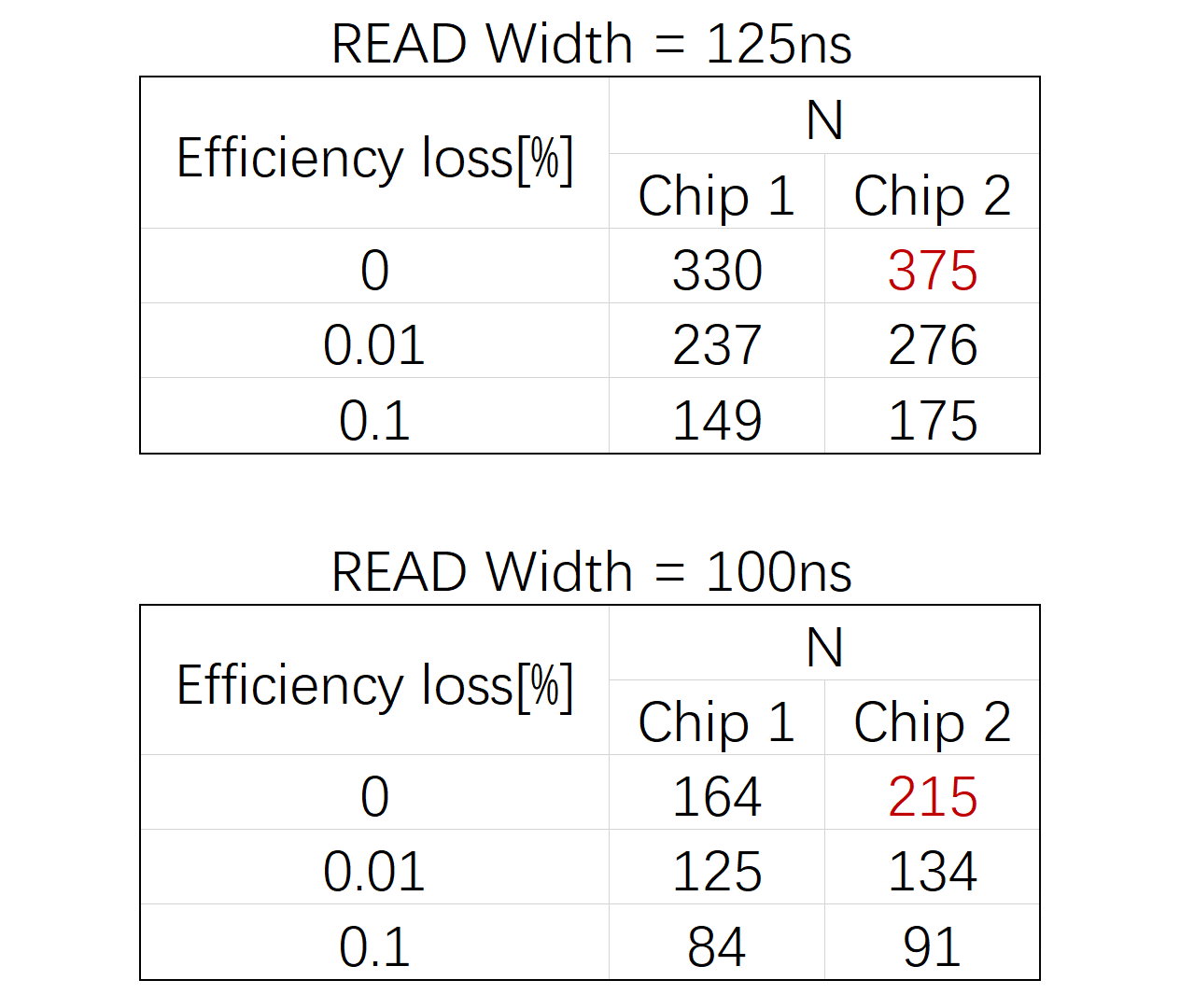}}
	\subfigure[]{
		\includegraphics[width=0.30\columnwidth]{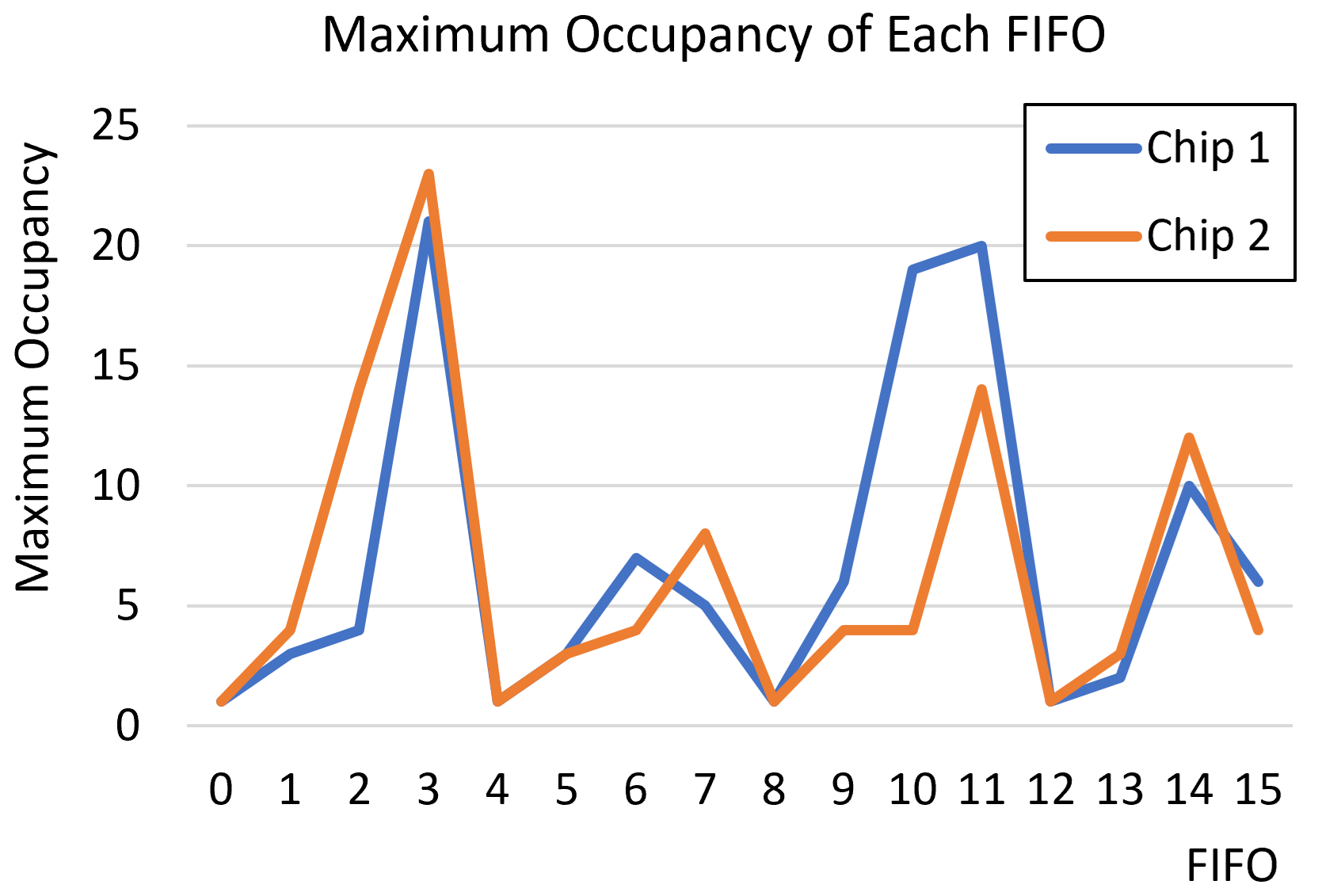}}
	\caption{(a) Distribution of latency from hit to data packet arrival at the multi-bank circular buffer. (b) Efficiency loss when reducing memory resources by truncating the long tail. (c) Maximum occupancy of 16 FIFOs at the frontend of the multi-bank circular buffer.}
	\label{FIG:3}
\end{figure*}
\subsection{Output link configuration and bandwidth utilization}
As shown in \textcolor{blue}{Figure 4}, the back end of the multi-bank circular buffer consists of an asynchronous FIFO, six finite state machines (FSMs) and six 8-bit to 1-bit serializers. To accommodate different hit densities, both the number and the frequency of the output serial links are configurable. The frequencies of the output serial links and the read clock of the asynchronous FIFO are configured together, maintaining a fixed 8:1 ratio to match the 8-bit to 1-bit serialization. The FSMs split long packets into individual bytes and send them to serializers. They are designed to enable back-to-back transmission during handshaking with the asynchronous FIFO, preventing bubbles and ensuring full bandwidth utilization. 

Simulation results for chip 1 show that when the Normal Compact format is used with six 1.28 Gbps output links, the bandwidth utilization of the six links is 99.5\%, 95.1\%, 88.9\%, 79.3\%, 63.5\% and 39.3\%, respectively, decreasing due to prioritization among the links. The maximum occupancy of the asynchronous FIFO is 4. For chip 2, the corresponding bandwidth utilization values are 99.8\%, 96.6\%, 93.4\%, 88.5\%, 80.8\% and 67.3\%, and the maximum FIFO occupancy is 6. The link provision is sufficient for the data volume in the highest-hit-density region.

\section{Conclusion and future work}
The behavioral-level simulation of COFFEE series chips shows that the column-drain readout mechanism operating under the high particle hit rate in LHCb Upgrade II demands a single readout cycle of no more than 100 ns to maintain detection efficiency. And shortening the single readout cycle will also significantly reduce the backend memory resource requirements and circuitry complexity.

In the peripheral readout architecture adapted to the BXID-sharing data format, the multi-bank circular buffer requires large memory resources due to extremely high particle hit rates in certain bunch crossings and queuing latency within pixels. The peripheral readout area is evaluate to be sufficient for the required memory resources in the advanced 55 nm process. However, most of the memory is idle most of the time, so there is room for optimization. More intelligent scheduling algorithms will be applied in the future.

It should be noted that in the simulation, the peripheral readout circuitry was decomposed into several stages. Each stage was assumed to operate under the ideal condition that the backend was always ready to receive data, meaning no back-pressure was applied. In future work, a global simulation will be conducted.
\section*{Acknowledgments}
This work was partially supported by the National Key Research and Development Program of China under Grant number 2023YFA1606300, the National Natural Science Foundation of China (NSFC) under Grant numbers W2443008, 11961141015, 12188102 and 12175245, the science and technology innovation project of Institute of High Energy Physics Chinese Academy of Sciences with the fund number 2024000077, and the High Energy Physics Research Center of Henan Academy of Sciences. 








\bibliographystyle{cas-model2-names} 

\bibliography{cas-refs}



\end{document}